\documentstyle[12pt]{article}
\def\ep{\epsilon}

\def\o{\over}

\def\ni       {\noindent}
\def\lb       {\left( }
\def\rb       {\right) }

\def\lbb     {\left[ }
\def\rbb      {\right] }
\def\comma      { \, , }
\def\period     { \, . }
\def\semiket#1  { \, #1 \, \rangle \, }
\def\del        {  \partial  }

\def\half       {  {1\over 2}  }
\def\abs#1      {  \, \vert #1 \vert \,   }
\def\Im#1    { \, {\rm Im } \, #1  }
\def\Re#1    { \, {\rm Re}  \, #1  }
\def\binom#1#2 { \vecii{ {}_{#1} }{\raisebox{.5ex}{$ {}^{#2} $}} }
\def\sqbinom#1#2 { \Bigl(\begin{array}{c} {}_{#1}
                       \\ \raisebox{.5ex}{${}^{#2}$} \end{array}\Bigr)^2  }

\def\calA    { {\cal A} }
\def\calB    { {\cal B} }
\def\calD    { {\cal D} }

\def\calF    { {\cal F} }
\def\calH    { {\cal H} }
\def\calV    { {\cal V} }
\def\calO    { {\cal O} }

\def\ybar     {\bar{y}}

\def\alphabar     {\bar{\alpha}}
\def\betabar     {\bar{\beta}}
\def\thetabar     {\bar{\theta}}

\def\Pbar   {\bar{P}}
\def\Fbar   {\bar{F}}
\def\Ebar   {\bar{E}}
\def\betabar  {\bar{\beta}}
\def\alphabar {\bar{\alpha}}

\def\mtil    {\tilde{m}}
\def\ntil    {\tilde{n}}
\def\atil    {\tilde{a}}
\def\btil    {\tilde{b}}
\def\ctil    {\tilde{c}}
\def\dtil    {\tilde{d}}

\def\gammabar  { \bar{\gamma} }

\def\3F2  { {}_3F_2 }
\def\2F1  { {}_2F_1 }
\def\G#1#2 { G{ #1 \brack #2} }
\def\F#1#2 { F{ #1 \brack #2} }
\def\vecii#1#2      {  \Bigl(\begin{array}{c}#1\\#2\end{array}\Bigr)  }
\def\veciii#1#2#3   {  \left(\begin{array}{c}#1\\#2\\#3\end{array}\right)  }
\def\matrixii#1#2#3#4            {  \Biggl( \begin{array}{cc}#1&#2\\#3&#4
                                       \end{array} \Biggr) }
\def\matrixiii#1#2#3#4#5#6#7#8#9 {  \left(\begin{array}{ccc}#1&#2&#3\\
                                     #4&#5&#6\\#7&#8&#9\end{array}\right)  }
\def\eqb         {  \begin{eqnarray}  }
\def\eqe           {  \end{eqnarray}  }
\def\nn               {  \nonumber  }
\def\sectionnumbering { \setcounter{equation}{0}
         \renewcommand{\theequation}{\arabic{section}.\arabic{equation}}}
\def\appendixnumbering { \setcounter{equation}{0}
         \renewcommand{\theequation}{\Alph{section}.\arabic{equation}}}
\def\mysection#1{ \addtocounter{section}{1} \setcounter{subsection}{0}
                 \sectionnumbering
   \par \bigskip
      \par \bigskip \noindent
   {\bf \arabic{section} \quad  #1 }
    \par \bigskip}
\def\appsection#1{\addtocounter{section}{1} \setcounter{subsection}{0}
                 \appendixnumbering
   \par \bigskip    
  \par \bigskip \noindent
   {\bf \Alph{section} \quad  #1 }
    \par \bigskip} 
\def\mysubsection#1{\addtocounter{subsection}{1}
      \par \bigskip \noindent  {\normalsize\it
      \arabic{section}.\arabic{subsection} \quad #1  }
   \par \medskip }

\def\csectionast#1    { \begin{center}
    {\large\bf #1  }   \end{center} \par \bigskip}
\def\titleandfile#1#2   {  \begin{center}{\large\bf #1}\end{center}
                            \par\begin{flushright} #2 \end{flushright}  }

\renewcommand{\thefootnote}{\fnsymbol{footnote}}

\newenvironment{namelist}[1]{%
   \begin{list}{}
      {
       \settowidth{\labelwidth}{#1}
       \setlength{\leftmargin}{1.1\labelwidth}}
}{%
 \end{list}}

\begin{document}
%
\def\papertitlepage{\baselineskip 3.5ex \thispagestyle{empty}}
\def\preprinumber#1#2#3{\hfill \begin{minipage}{4.2cm}  #1
              \par\noindent #2
              \par\noindent #3
             \end{minipage}}
\renewcommand{\thefootnote}{\fnsymbol{footnote}}
%
%
\papertitlepage
\setcounter{page}{0}
\preprinumber{KEK Preprint 2002-89}{UTHEP-459}{hep-th/0209043}
\baselineskip 0.8cm
\vspace*{2.0cm}
\begin{center}
{\large\bf Penrose limits and Green-Schwarz strings}
\end{center}
\vskip 4ex
\baselineskip 1.0cm
\begin{center}
        { Shun'ya~ Mizoguchi\footnote[2]{\tt mizoguch@post.kek.jp} } \\
 \vskip -1ex
    {\it High Energy Accelerator Research Organization (KEK)} \\
 \vskip -2ex
    {\it Tsukuba, Ibaraki 305-0801, Japan} \\
 \vskip 2ex   
 {Takeshi~ Mogami\footnote[3]{\tt mogami@het.ph.tsukuba.ac.jp}}
        \ \ {and} \
        { Yuji  ~Satoh\footnote[4]{\tt ysatoh@het.ph.tsukuba.ac.jp}}  \\
 \vskip -1ex
    {\it Institute of Physics, University of Tsukuba} \\
 \vskip -2ex
   {\it Tsukuba, Ibaraki 305-8571, Japan}

\end{center}
\vskip 10ex
%
\baselineskip=3.5ex
\begin{center} {\large\bf Abstract} \end{center}

We discuss the Green-Schwarz action for type IIB strings in general
plane-wave backgrounds obtained as Penrose limits from
any IIB supergravity solutions with vanishing background fermions.
Using the normal-coordinate expansion  in superspace,
we prove that  the light-cone action is necessarily quadratic
in the fermionic coordinates.
This proof is valid for more general pp-wave backgrounds under
certain conditions. We also write down the complete quadratic 
action for general bosonic on-shell backgrounds in a form in which 
its geometrical meaning is manifest both in the Einstein and 
string frames. When the dilaton and 1-form field strength are vanishing, 
and the other field strengths are constant, our string-frame action
reduces, up to conventions, to the one which has been written down 
using the supercovariant derivative.

\vskip 2ex
%
%
%
%
%
\vspace*{\fill}
\ni
September 2002
\newpage
\renewcommand{\thefootnote}{\arabic{footnote}}
\setcounter{footnote}{0}
\setcounter{section}{0}
\baselineskip = 0.6cm
\pagestyle{plain}
%
%
%
\mysection{Introduction}
In general relativity, any solution leads to a plane-wave solution
in the Penrose limit \cite{Penrose}. This is also generalized to any
supergravity theories in ten and eleven dimensions \cite{Gueven}.
The maximally supersymmetric solution supported by a constant
RR flux \cite{Blau-FHP} is a particular example in IIB supergravity,
which is realized as a Penrose limit of $AdS_5 \times S_5$
\cite{Blau-FHP2}. Remarkably, the string theory on this background
is exactly solvable in the Green-Schwarz formalism \cite{Metsaev}.
Moreover, it provides a model for analyzing the AdS/CFT correspondence
in the stringy regime \cite{Berenstein-MN}.
The string theory on the plane wave thus opened up a way
to study important questions in string theory such as
strings in curved spacetime, strings in non-trivial RR backgrounds and
stringy aspects of the AdS/CFT correspondence.

After the works of \cite{Blau-FHP,Metsaev,Berenstein-MN},
string models on other plane-wave backgrounds have also been studied.
For a partial list, see \cite{Russo-T}-\cite{Berkovits-M}.
In analyzing those models, it is important to know whether or not
the Green-Schwarz string action in light-cone gauge is quadratic
in the fermionic coordinates. In the original case \cite{Metsaev},
this was directly shown by the supercoset method \cite{Metsaev-T},
which uses symmetry superalgebras associated with backgrounds.
When the plane wave is obtained as a Penrose limit (and dimensional
reduction) from $AdS\times M$ spaces in ten or eleven dimensions,
this special property is also shown \cite{Metsaev-T2,Russo-T,IIA}
by using the Green-Schwarz string or membrane actions
for the corresponding $AdS\times M$ spaces \cite{Metsaev-T}-\cite{Membrane}.
However, there is no proof for general plane-wave backgrounds.

For general bosonic curved backgrounds which satisfy supergravity
equations of motion, the Green-Schwarz actions are known 
up to quadratic order in the fermionic coordinates:
In the type IIA and IIB cases, they were derived in \cite{Cvetic-LPS} by
starting from the corresponding membrane action in eleven dimensions.
In the heterotic case, it was also derived in \cite{Atick-D}
by starting from the $\kappa$-symmetric action
written in superfields \cite{Witten,Atick-DR} and using
the `$\theta$-expansion' \cite{Atick-D,Mukhi}, which is a generalization
of the ordinary
normal-coordinate expansion for bosonic non-linear $\sigma$-models.
The $\theta$-expansion was also applied to the IIB case
in \cite{Gates-MRV,Sahakian}.
For constant axion, $3$- and $5$-form field
strengths and vanishing dilaton, the IIB action was written down
in a concise form in terms of the supercovariant 
derivative \cite{Metsaev-T2}.
Once a Green-Schwarz action in a background is assured to terminate
at quadratic order in fermions, one can use those results.
What is a little worrying here is that there are issues on signs,
factors and normalizations \cite{Corrado-HKW}. 
Thus, it would be of some use to provide an explicit derivation of 
the quadratic actions in an independent manner, in which their geometrical 
meaning also becomes clear.
  
In this paper, we discuss the Green-Schwarz action for type IIB strings
in general plane-wave backgrounds obtained as Penrose limits of
any IIB supergravity solutions with vanishing background fermions.
(These backgrounds always possess Killing spinors, at least locally.)
We start from the $\kappa$-symmetric action for general on-shell
supergravity backgrounds written in terms of superfields \cite{Grisaru-HMNT},
and use the $\theta$-expansion. This method has an advantage that
it can be applied to any backgrounds. On the other hand, it also has
a disadvantage that, in general,  it is difficult to carry out the expansion
and express the result in terms of physical fields to higher orders.
Because of this, the known complete IIB Green-Schwarz actions
in non-trivial backgrounds have been derived by
the supercoset method, which relies on high symmetries
of backgrounds \cite{Metsaev},\cite{Metsaev-T}-\cite{AdS3},
or by taking the Penrose limit for these known actions
\cite{Berenstein-MN,Metsaev-T2,Russo-T} 
(see, also \cite{Maldacena-M,Berkovits-M}).
In spite of that, we show that the $\theta$-expansion
is effective enough, for the plane-wave solutions we consider,
to prove that the Green-Schwarz action in  light-cone gauge
is quadratic in the fermionic coordinates.
The proof is also valid under certain conditions
for more general pp-wave backgrounds,
of which  the plane-wave solutions we consider are a subclass.
By using this systematic expansion method,
we also write down the complete quadratic action for general bosonic
on-shell backgrounds in a form in which its geometrical
meaning becomes clear both in the Einstein and string frames.
We confirm that the quadratic action in the string frame is expressed
by the supercovariant derivative.
We note that any of the bosonic background fields do not have to be constant
in our discussion. When the dilaton and 1-form field strength are
vanishing, and the other field strengths are constant,
we find an agreement with the action in \cite{Metsaev-T2} 
after taking into account the differences of conventions.

The organization of this paper is as follows. In section 2,
we give a brief summary about the Penrose limit, the superspace
formulation of IIB supergravity, the IIB Green-Schwarz
action for general on-shell backgrounds and the $\theta$-expansion.
In section 3, we prove that the light-cone action which we consider is
quadratic in the fermionic coordinates. In section 4,
we write down the quadratic action both in the Einstein and string fames.
We conclude with a discussion
in section 5. Our notation and conventions are summarized in Appendix A.
Some components of the superfields, which are needed in our discussion,
are listed in Appendix B.

\vspace{6ex}
\ni
{\it Note added} 

After this work was essentially completed, a preprint \cite{Russo-T2}
appeared in which it is proved that the IIB Green-Schwarz action
is quadratic in the fermionic coordinates for a general class of 
pp-wave solutions which have non-constant 3- and 5-form field strengths, 
and vanishing dilaton, 1-form field strength and background fermions. 
The method, however, is different: In \cite{Russo-T2}, 
the general possible terms in the Green-Schwarz action are 
listed by examining their tensor structure,
while in our $\theta$-expansion each term is automatically obtained 
explicitly with precise coefficients. The $\theta$-expansion method
can also be applied to more general backgrounds other than pp-waves, 
to show that the corresponding Green-Schwarz action terminates 
at some order, and obtain its explicit form.
In addition, a revised version of \cite{Cvetic-LPS} appeared
in the electronic archive, in which typos are corrected and explanations
of their conventions are added. 

\newpage
%
%
\mysection{Green-Schwarz action in general curved backgrounds}
\mysubsection{Penrose limits}
Plane-wave solutions obtained from any other solutions
in the Penrose limit have
a universal form. In the IIB case, which we are interested in,
the bosonic fields in the harmonic coordinate system are given by
\cite{Gueven}
\eqb
  ds^2 & = & dx^+dx^- - h_{\mtil \ntil}(x^+) x^{\mtil} x^{\ntil} (dx^+)^2
   - \delta_{\mtil\ntil} dx^{\mtil} dx^{\ntil}
   \comma \nn \\
   \tau  &=& \tau(x^+) \comma
    \label{pwsol}   \\
    \tilde{F}_{p} &=& \frac{1}{(p-1)!}
   \tilde{F}_{+ {\mtil}_1...{\mtil}_{p-1}}(x^+)
   dx^+ \wedge dx^{\mtil_1} \wedge...\wedge dx^{\mtil_{p-1}}
  \period \nn
\eqe
Here, $\mtil,\ntil = 1,2,...,8$, $x^\pm = x^0 \pm x^9$ and $p=3,5$.
We reserve un-tilted $m,n...$ to denote all spacetime directions;
$m,n = 0,1,...,9$. $h_{{\mtil}{\ntil}}$ are some functions,
$\tau$ is the complex scalar, and
$\tilde{F}_3$ and $\tilde{F}_5$ are the complex 3-form
and self-dual 5-form field strengths,
respectively. The corresponding $(p-1)$-form potentials $A_{p-1}$ can
have non-trivial components $A_{{\mtil}_1...{\mtil}_{p-1}}$ and
$A_{+{\mtil}_1...{\mtil}_{p-2}}$. The non-zero components
of the Riemann tensor $R_{klmn}$ and the Ricci tensor $R'_{mn}$
are only $R_{+{\mtil}+{\ntil}}$ and $R'_{++}$,
up to symmetry of the tensors.
(We put the prime for the Ricci tensor
to avoid confusion with the curvature 2-form appearing later.)
In addition, 
the non-vanishing Christoffel symbols $\Gamma^l_{mn}$ are
$\Gamma^{\mtil}_{++}, \Gamma^-_{+\mtil}, \Gamma^-_{++}$.
Since all the above fields depend only on $x^+$, the covariant
derivatives of the tensors, e.g., $ \nabla_k R'_{mn}$, are non-vanishing
only for $\nabla_+$. 

Similar properties hold also for tensors with tangent-space indices.
This is confirmed by noting that the spacetime indices are converted
to the tangent-space indices through the vielbein, which are read off from
\eqb
  e^+ = dx^+ \comma \quad
  e^- = dx^- - h_{\mtil\ntil} x^{\mtil}x^{\ntil} dx^+ \comma \quad
  e^{\atil} = dx^{\mtil} \delta_{\mtil}^{\ \atil}
  \period
\eqe
Here, we denote the tangent-space indices by $a,b = 0,1,...,9$, and
$\atil,\btil = 1,...,8$. The non-vanishing components of the
spin connection $\omega_{m, a}^{\ \ \ \  b}$ are only
$\omega_{+, \atil}^{\ \ \ \ -}$. 
We also denote the covariant derivative 
for tensors with the tangent-space indices by $D_a$.

{}From these observations, we find that
\begin{namelist}{(III)}
  \item[(I)] The non-vanishing components of the tensors (tensor densities)
   have neither lower `$-$' indices nor upper `$+$' indices, except
 for $g_{mn}, \ep_{m_1,..., m_{10}}$,
   $\eta_{ab}, \ep_{a_1,..., a_{10}}$.
 \item[(II)] The tensors have lower `$+$' indices at least as many as
     their mass dimension when all the indices are lowered.
 \item[(III)] The number of the lower `$+$' plus upper `$-$' is
  preserved when the indices are raised, lowered, or contracted with
 indices of other tensors.
\end{namelist}
Here, for example, $R_{klmn}$ has dimension two, and $\tilde{F}_p$ has
one. 

\mysubsection{IIB Green-Schwarz action in curved backgrounds}
The $\kappa$-symmetric IIB Green-Schwarz superstring action
was constructed in \cite{Grisaru-HMNT} for general curved backgrounds
satisfying the supergravity constraints and, hence, the supergravity
equations of motion. To write it down, we need
the superspace formulation of type IIB supergravity \cite{Howe-W}.
In the following, we follow the notation and conventions in \cite{Howe-W}
unless otherwise stated.

The superspace coordinates are then denoted by
$z^M =  (x^m,\theta^\mu,\theta^{\bar{\mu}})$. The field content of
the IIB theory is as
follows: the frame 1-forms
$E^A =  (E^a, E^\alpha, E^{\alphabar} = \bar{E}^\alpha) =
 dz^M E_M^{\ \ A} $,
the $SO(1,9) \times U(1)$ connection 1-form $\hat{\Omega}_A^{\ \ B}$;
the complex 2-form potential $\calA$ and the real 4-form potential $B$.
From these fields, the torsion 2-form $T^A$, the curvature 2-from
$R_A^{\ \ B}$, the complex 3-form field strength $\calF$, and the 5-form
field strength $Z$ are constructed. In addition, there appears
a scalar superfield, which is an element of $SU(1,1)$,
\eqb
   \calV &=& \matrixii{u}{v}{\bar{v}}{\bar{u}} \comma \qquad
    u \bar{u} - v \bar{v} = 1 \period
\eqe  
From this, one obtains the 1-form,
\eqb
  \calV^{-1} d\calV = \matrixii{2iQ}{P}{\Pbar}{-2iQ}
  \comma \quad Q = \bar{Q} \period
\eqe 
The real 1-form $Q$ is identified with the $U(1)$ part of
$\hat{\Omega}_A^{\ \ B}$,
whereas the complex 1-form $P$ is expanded as
$P = E^a P_a + E^\alpha P_\alpha - E^{\alphabar} P_{\alphabar}$ with
\eqb
  P_\alpha = -2 \Lambda_\alpha \comma \quad P_{\alphabar} = 0
  \period \label{Palpha}
\eqe 
$\Lambda_\alpha$ contains the physical spin $1/2$ field of the theory.
Also, the scalar field $\calV$ and the complex 3-form $\calF $ combine
to form an $SU(1,1)$ invariant 3-form field strength $F$:
\eqb
   (\bar{\calF}, {\calF}) \calV &=& (\Fbar,F)
  \period
\eqe 
These fields in the IIB theory satisfy superspace constraints and
Bianchi identities. We list their relevant components in Appendix B.
 
By making use of the above formulation,
the IIB Green-Schwarz action for general on-shell curved backgrounds is
given by 
\eqb
   I &=& \half \int d^2 \xi \,
   \lbb \sqrt{-g} g^{ij} \Phi E_i^{a} E_j^{ b}
         \eta_{ab} + \epsilon^{ij} E^{ B}_i E^{ A}_j \calB_{AB} \rbb
   \period \label{GSaction}
\eqe
Here, $g^{ij}$ is the world-sheet metric, $\epsilon^{ij}$ is
the anti-symmetric tensor density and
$E_i^{ A} \equiv \del_i z^M E_M^{\ \ A}$. $\calB_{AB}$ is related to the
real closed 3-form $\calH$ by
\eqb
  \calH \equiv \calF + \bar{\calF} = d \calB
  \period
\eqe
$\Phi$ is given by the components in $\calV$:
\eqb 
   \Phi = w = \bar{w} \comma \qquad w = u-\bar{v}
   \comma \label{gauge}
\eqe 
where we have taken a specific local $U(1)$ gauge so that  $w$ becomes real.
In this gauge, one can derive
\eqb
   D_A \Phi &=& -\half \Phi (P_A + \Pbar_A) \comma \nn \\
     Q_A & = & \frac{i}{4} (P_A - \Pbar_A) \comma \label{PQ} \\
  \calH &=& \Phi (F+\Fbar) \period \nn
\eqe
 
When the background satisfies the supergravity constraints,
the action (\ref{GSaction}) has the $\kappa$
symmetry given by
\eqb
   \delta E^a  &\equiv & \delta z^M E_M^{\ \ a} \ = \ 0 \comma \nn \\
   \delta E^\alpha  &\equiv & \delta z^M E_M^{\ \ \alpha}
      \ = \ 2 E_i^{ a} (\sigma_a)^{\alpha \beta} g^{ij} \eta_{j \beta}
   \comma \nn \\ 
   \delta \Ebar^{\alpha}  & = &  \overline{(\delta E^\alpha)} \comma  \nn\\
  \delta \Phi &=& \Phi (\delta E^\alpha \Lambda_\alpha
       - \delta \Ebar^\alpha \bar{\Lambda}_\alpha ) \comma \\
   \delta (\sqrt{-g} g^{ij}) &=&
      4i (g^{ik} \ep^{jl} + \ep^{ik} g^{jl})
     (E^{\alpha}_k \eta_{l \alpha} + \Ebar^{\alpha}_k
   \bar{\eta}_{l \alpha})
     \nn \\
  && \qquad  
    + \, 2 (g^{ij}\ep^{kl} -2\ep^{kj}g^{il}) E^{ c}_{k}
     (\sigma_c)^{\alpha\beta}
    (\bar{\eta}_{l\alpha}\Lambda_\beta -\eta_{l\alpha}\bar{\Lambda}_\beta)
   \period \nn
\eqe
Here, $(\sigma_a)^{\alpha\beta}$ (and $(\sigma_a)_{\alpha\beta}$) are
the $16 \times 16$ $\gamma$-matrices. $ \eta_{j\beta} $ satisfies
$ \sqrt{-g} g^{ij} \eta_{j \beta} = - \ep^{ij}  \bar{\eta}_{j\beta}$.
Note that the action  in (\ref{GSaction}) is expressed 
in the Einstein frame.
\mysubsection{Expansion by fermionic coordinates}
In order to analyze a model described by
the Green-Schwarz action in the previous subsection, it is desirable
to express it in terms of the superspace coordinates $z^M$.
A systematic method for such a purpose was developed 
in the heterotic case ($\theta$-expansion) \cite{Atick-D},
following the algorithm in \cite{Mukhi}.
This is a sophisticated version of the ordinary normal-coordinate
expansion for
bosonic non-linear $\sigma$-models. It is also straightforward to apply
this method to the IIB case \cite{Gates-MRV,Sahakian}.

Following this method, the action $I$ in (\ref{GSaction})
is covariantly expanded
as a functional of $z^M$. Let $y^M$ be a tangent vector of a geodesic
through $z^M$, which satisfies $ y^B D_B y^A = 0 $ with
$y^A = y^M E_M^{\ \ A}$. We then have
\eqb
   I (z') &=& e^{\Delta(z,y)} I(z)
  \comma \label{thetaex} \\
    \Delta (z,y) & =  &\int d^2 \xi  \, y^A(\xi) D_A(\xi) \period \nn
\eqe
Here, $z^{'M} = z^M + y^M $ when the normal coordinates are taken.
$D_A(\xi)$ is the functional covariant derivative. For example,
for a supervector $X^B(\xi)$,
\eqb
  D_A(\xi) X^B(\xi') = E_A^{\ \ M} \lb z(\xi) \rb
  \frac{\delta X^B(\xi')}{\delta z^M(\xi)}
  + \delta(\xi,\xi') (-1)^{[C][A]} X^C(\xi)
  \hat{\Omega}_{A, C}^{\ \ \ \, B} (\xi)
  \comma
\eqe
where $\delta(\xi,\xi')$ is the delta function and $[A]$ takes $0$ for
bosonic indices and $1$ for fermionic indices.
Since we are considering the IIB case, the covariant derivative
includes the $U(1)$
connection as well as the ordinary spin connection.
From the definition of $\Delta$, it follows that
\eqb
  \Delta X^{BC...}_{DE...} &=& y^A D_A X^{BC...}_{DE...} \comma \nn \\
  \Delta y^A &=& 0 \comma \label{actionDelta} \\
  \Delta E_i^{ A} &=& D_i y^A + E_i^{C} y^B T_{BC}^{\ \ \ A} \comma \nn \\
  \Delta (D_i y^A) &=& y^B E_i^{ D} y^C R_{CDB}^{\ \ \ \ \ A}
   \comma  \nn
\eqe
with $X^{BC...}_{DE...}$ an arbitrary tensor and
$D_i \equiv E_i^{ A} D_A$.

Since we would like to obtain the expansion with respect to the fermionic
variables only, at the final stage of the expansion, we set
\eqb
   z^{M} = z_0^M \equiv (x^m,0,0) \comma &&
   y^M =  y_0^M \equiv (0, y_0^\mu, y_0^{\bar{\mu}} )
   \comma \label{z0}
\eqe 
 At $z^M = z^M_0$, the superfields retain their lowest components
in the fermionic coordinates:
\eqb
  && E_m^{\ a} = e_m^{\ a} (x) \comma \quad
  E_m^{\ \alpha}   =  \psi_m^{\ \alpha} (x) \comma \quad
  E_m^{\ \bar{\alpha}}  =  \bar{\psi}_m^{\ \alpha} (x) \comma \quad
  E_\mu^{\ \alpha}  =   \delta_\mu^{\ \alpha}   \comma \quad
  E_{\bar{\mu}}^{\ \bar{\alpha}}  =  - \delta_{\bar{\mu}}^{\ \bar{\alpha}}
  \comma \nn \\
  && E_\mu^{\  a}  =  E_{\bar{\mu}}^{\  a} =
    E_{{\mu}}^{\ \bar{\alpha}} =  E_{\bar{\mu}}^{\ {\alpha}}
   = 0 \comma \nn  \\
  && Q_m = q_m (x) \comma \label{EOmega0} \\
 &&  \hat{\Omega}_{m, a}^{\ \ \ \ b}
  = \omega_{m, a}^{\ \ \ \ b} (x)\comma \quad
  \hat{\Omega}_{m, \alpha}^{\ \ \ \ \beta}  =
     \frac{1}{4}  (\sigma^{ab})_\alpha^{\ \beta} \omega_{m, ab}(x)
    + i \delta_\alpha^{\ \beta} e q_m(x) \comma \nn \\
 &&  \hat{\Omega}_{\mu, A}^{\ \ \  \, B} =
  \hat{\Omega}_{\bar{\mu}, A}^{\ \ \  \, B} \ = \ 0
  \period \nn 
\eqe
Here,  $ e $ represents  the charge of the fields, e.g., $e = +1$ for
$\theta^\alpha$. Thus, for $y_0^A \equiv y_0^M E_M^{\ \ A}(z_0)$,
one has
\eqb
   y_0^a &=& 0 \comma \quad
  y_0^\alpha \ = \ y_0^\mu E_\mu^{\ \alpha} \ \equiv \ \theta^\alpha
   \comma \quad 
  y_0^{\bar{\alpha}} \ = \ -y_0^{\bar{\mu}}
   E_{\bar{\mu}}^{\ \bar{\alpha}} \ \equiv \  \thetabar^\alpha
 \period \label{yA0}
\eqe

Since the terms dropped in setting $z^M = z_0^M$ may give contributions 
at higher orders by further acting with $\Delta$,
 (\ref{z0})-(\ref{yA0}) cannot be substituted
at intermediate stages of the expansion. However, there is an exception:
Let us recall the rules in (\ref{actionDelta}) and the fact that
tensors with the Lorentz indices such as $R_A^{\ \ B}$ are non-vanishing
only when both of $A,B$ are bosonic or fermionic. One then finds
that acting with $\Delta$ on $y^a$ yields no contribution even
at higher orders and, thus, $y^a =0$ can be set at any stage.
We also note that, in the case we are interested in,
the background fermions are set to be zero at the final stage.
We express this by an arrow: $ W \to W'$ means
$W$ (background fermions = 0) $= W'$.

In this way, one can in principle
derive the IIB Green-Schwarz action for general on-shell
backgrounds explicitly in terms of
$(x^m,\theta^\alpha, \theta^{\bar{\alpha}})$. The computation
at low orders is straightforward.
For example, the zero-th order term in the expansion, $I^{(0)}$, is obtained
simply by setting $z^M = z_0^M$ in $I(z)$.
At the first order $I^{(1)} = \Delta I$, we need to compute
$\Delta E_i^{a}$ and $\Delta \Phi$. By using
(\ref{actionDelta}), (\ref{PQ}), and components in (\ref{Palpha})
and Appendix B, they are found to be
\eqb
  \Delta \Phi &=& \Phi (y^\alpha \Lambda_\alpha
   - \ybar^\alpha \bar{\Lambda}_\alpha) \comma \nn \\
  \Delta E_i^a & = &
   -i [y^\alpha (\sigma_a)_{\alpha\beta} \Ebar_i^\beta
   + \ybar^\alpha (\sigma_a)_{\alpha\beta} E_i^\beta ]
   \period \label{DeltaEa}
\eqe
We then have
\eqb
  L^{(1)} & = &   
    \half \lbb \sqrt{-g} g^{ij} \lb
     \Delta \Phi E_i^a E_j^b  + 2 \Phi E_i^{ a} \Delta E_j^{ b} \rb
         \eta_{ab} + \epsilon^{ij} y^C E^{ B}_i E^{ A}_j \calH_{ABC} \rbb
    \nn  \\
   &=& \half \Phi E_i^{ a}
    ( \sqrt{-g} g^{ij} y^\alpha
   - \ep^{ij} \ybar^\alpha ) (\sigma_a)_{\alpha\beta}
     \lbb (\sigma_b)^{\beta\gamma} \Lambda_\gamma  E_j^{ b}
    - 2i\Ebar_j^{ \beta}
   \rbb  + {\rm h.c.} \
 \comma \label{L1} 
\eqe
where $I^{(n)} = \int d^2\xi \, L^{(n)} $.
{h.c.} stands for hermitian conjugates.

After some amount of algebras, the expression at the second order is
also obtained. We discuss this in section 4.
However, as mentioned in Introduction, at higher orders it is becoming
more and more difficult to carry out the expansion and express the results
in terms of physical fields.

\mysection{Light-cone action is quadratic in fermionic coordinates}
In this section, we prove that the light-cone action
in the plane-wave backgrounds summarized in subsection 2.1 is quadratic in
the fermionic coordinates $\theta,\thetabar$.
We also see that the proof is valid for more general pp-wave backgrounds
under certain conditions.

Before starting our discussion, we would like to make
comments on the works in \cite{Atick-D} and \cite{Sahakian}.
In \cite{Atick-D}, it was argued that the light-cone
Green-Schwarz action for the heterotic theory in a certain class
of curved backgrounds is quadratic in the fermionic coordinates.
The argument was based on the claim,
   $O^{-\atil\btil} O^{-\ctil\dtil} = 0 $
 for a Majorana-Weyl spinor in light-cone gauge
 $(\sigma^+)_{\alpha\beta} \vartheta^\beta = 0 $, where
 $O^{-\atil\btil} =
   \vartheta^\alpha (\sigma^{-\atil\btil})_{\alpha\beta} \vartheta^\beta$.
However, this cannot be true:
If this was true, one could show that any term of the form
$\vartheta^{\alpha} \vartheta^{\beta} \vartheta^{\gamma} \vartheta^{\delta}$
would be vanishing by making use of the Fierz transformations.
In fact, we have checked that
the numbers of the independent non-vanishing terms
of $\vartheta^{\alpha} \vartheta^{\beta} \vartheta^{\gamma}
\vartheta^{\delta}$ and those of $O^{-\atil\btil} O^{-\ctil\dtil}$
are the same. Similar confusions are found in \cite{Sahakian}
in the argument that some quartic terms in fermions are vanishing.
In the following discussion, we do not use this kind of argument.

As a preliminary to our proof, let us note several facts.
First, in light-cone gauge for the IIB case,
\eqb
   (\sigma^+)_{\alpha\beta} \theta^\beta =
   (\sigma^+)_{\alpha\beta} \thetabar^\beta = 0
  \period \label{lightcone}
\eqe 
Furthermore, at $z^M = z_0^M$, one has (\ref{EOmega0}) and
\eqb
  D_a \theta^\alpha  &=&
   \del_a \theta^\alpha
   - \frac{1}{4} \omega_{a,bc}(\sigma^{bc})^\alpha_{\ \beta} \theta^\beta
   + i q_a \theta^\alpha \period \label{Dtheta}
\eqe
From these and (I) listed in section 2, it follows that
\eqb
  (\sigma^+)_{\alpha\beta} D_i \theta^\beta =
   (\sigma^+)_{\alpha\beta} D_i \thetabar^\beta = 0
  \period
\eqe
In comparing the connections in (\ref{Dtheta}) and (\ref{EOmega0}), 
we note that $(\sigma^{bc})^\alpha_{\ \beta}
= - (\sigma^{bc})^{\ \, \alpha}_{\beta}$.
(The expression of the spin connection in terms of the vielbein
is read off from that of the torsion in \cite{Howe-W}.)
    
Second, since we are considering the case in which
background fermions are vanishing, background fields with odd number of
spinor indices are vanishing. From the representation theory
of the Lorentz group, background fields with even number of spinor indices
should be expressed by terms of the form:
($\gamma$-matrices)$\times$(background fields only with bosonic indices).
Thus, the spinors $\theta, \thetabar$ have to be contracted
with the $\gamma$-matrices in order to obtain non-vanishing terms including
them. Since $\theta, \thetabar$ has a definite chirality,
the possible contractions in light-cone gauge give only the following
bilinears, 
\eqb
  \thetabar \sigma^- \theta \comma \quad
  \theta \sigma^{-\atil\btil} \theta \comma \quad
  \thetabar \sigma^{-\atil\btil} \thetabar \comma \quad
  \thetabar \sigma^{-\atil\btil} \theta \comma \quad
  \thetabar \sigma^{-\atil\btil\ctil\dtil} \theta \period
  \label{bilinear}
\eqe

Third, recalling (\ref{L1}) and (\ref{actionDelta}),
one finds that the fields appearing in $I^{(n)}$ ($n\geq 1$) are: (1)
$\Phi, H_{ABC}, T_{AB}^{\ \ \  C}, R_{ABC}^{\ \ \ \ \ D}$,
 (their covariant derivatives);
(2) $\eta_{ab}, \epsilon_{a_1, ..., a_{10}}$;
 (3) $E_i^A, y^{\hat{\alpha}},
D_i y^{\hat{\alpha}}$ with $\hat{\alpha} = (\alpha, \alphabar)$.
The 2-form potential, for example, does not appear
in the expansion except at the zero-th order. It could appear only
through the decomposition of tensors with two spinor indices of the form
$X_{\alpha\beta}$. The decomposition, however,  is
carried out algebraically by using the constraints and Bianchi identities,
but they do not include the 2-form potential.
After setting $z^M = z_0^M$ and
background fermions to be zero, the background fields in set (1)
are expressed by the bosonic physical fields described in subsection 2.1.
Otherwise, the superspace would contain degrees of freedom which are
redundant for the IIB theory. Thus, the fields which appear in the expansion
satisfy the properties (I)-(III) listed in section 2. In fact,
for the fields coming from (1), the property (II) becomes more precise:
\begin{namelist}{(II') II}
  \item[ (II') ]
 They have lower `$+$' indices  as many as
     their mass dimension when all the indices are lowered.
\end{namelist}

Now, let $W^{(2n;p)}_{a_1,...a_k}(x,\theta,\thetabar)$ with
mass dimension $p \geq 0$ be a term
which is made of $2n$-spinors and
the background bosonic fields in set (1) and (2) above.
Then, the following statement holds:
\begin{namelist}{IIIIIIII}
  \item[\rm Lemma: ]
    {\it In light-cone gauge, $W^{(2n;p)}_{a_1,...,a_k } \neq 0$
     only when $ m_+ = m_- + 2n + p$,
    where $m_+(m_-)$ is the number of $+(-)$ indices among
   $(a_1, ..., a_k)$.}
\end{namelist}
We show this by induction with respect to $n$. For $n=0$,
$W^{(0;p)}_{a_1,...,a_k }$ contains only bosonic fields in (1) and (2).
Since only $\eta_{-+}, \ep_{-+....}$ have lower $-$ indices, each
lower $-$ is accompanied by one lower $+$. Also,
$W^{(0;p)}_{a_1,...,a_k }$ has dimension $p$ and, hence,
has to have $p$ lower $+$ in addition to $+$'s in $\eta_{-+}, \ep_{-+....}$.
Thus, one can make $W^{(0;p)}_{a_1,...,a_k }$ only when $m_+ = m_- + p$.
We then suppose that the above statement holds for
$ W^{(2(n-1);p)}_{a_1,...,a_k }$. We also denote bilinears in
(\ref{bilinear})
by $O^{ \{ -r \} }$ with $ \{ -r \} $ expressing
their index structures:
$ \{ -r \} = -, - [\atil\btil], - [ \atil\btil\ctil\dtil ] $.
The brackets stand for anti-symmetrization.
Extracting one bilinear, $ W^{(2n;p)}_{a_1,...,a_k }$ takes the form
$ W^{(2n;p)}_{a_1,...,a_k }
 = \sum O^{ \{ -r \} } W^{(2(n-1);p+1)}_{ \{ - r \} a_1,...,a_k }$.
The right-hand side is non-vanishing only when
$ m_+ = (m_- +1) + 2(n-1) + (p+1) = m_- + 2n + p$, and so is
$W^{(2n;p)}_{a_1,...,a_k }$.

Using this result, we can analyze the possible order in $\theta, \thetabar$
of the action $I$. 
Taking into account (\ref{L1}) and the fact that
we eventually set the background fermions to be zero,
the possibly non-vanishing terms
in the $\theta$-expansion are of the forms, 
\eqb
   & (\Delta^{2k} \Phi)(\Delta^{2l} E_i^a)(\Delta^{2m} E_j^b)
   [ \theta (\sigma_{ab} + \eta_{ab}) (\Delta^{2n-1}\Lambda) ] \comma &
   \nn \\
   & (\Delta^{2k} \Phi)(\Delta^{2l} E_i^a) \theta^\alpha
   (\sigma_a)_{\alpha \beta} (\Delta^{2m-1} E_j^{\beta})
    \comma & \label{higerterm}
\eqe  
their hermitian conjugates and similar expressions obtained by replacing
$\theta$ with $\thetabar$.

$ \Delta^{2n} \Phi $ has dimension zero and $2n$ spinors.
Thus, the lemma shows that in light-cone gauge
this is non-vanishing only when $n=0$
after setting $z^M=z_0^M$ and the background fermions to be zero.
Similarly, $  \theta (\sigma_{ab} + \eta_{ab}) (\Delta^{2n-1}\Lambda)  $
has dimension 0 and $2n$ $(n \geq 1)$ spinors and, thus, this is
vanishing.

To know the possible order of
the remaining terms of the forms $\Delta^{2n} E_i^{ a}$ and
$\Delta^{2n-1} E_i^{ \alpha}$,
we need a little more analysis, because  these have an 
extra world-sheet index. Similarly to the heterotic case discussed 
in \cite{Atick-D},
the general form of $\Delta^{2n} E_i^a$
follows from (\ref{actionDelta}) to be given by
\eqb
  \Delta^{2n} E_{i a} & \to & W_{ab} E_i^b
   + W_{a \hat{\beta}} D_i \theta^{\hat{\beta}}
  \comma 
\eqe 
where $\theta^{\bar{\beta}} = \thetabar^{\beta}$. In our setting,
$W_{a \hat{\beta}}$ should be
\eqb
  W_{a\hat{\beta}} &\to& \theta^{\hat{\alpha}} W_{a \hat{\alpha}
\hat{\beta}}
  \comma 
\eqe
and $ W_{a \hat{\alpha} \hat{\beta}} $ takes the form for
$(\hat{\alpha},\hat{\beta}) = (\alpha, \beta)$
\eqb
  S_{ab} (\sigma^{b})_{\alpha\beta} + M_{abcd} (\sigma^{bcd})_{\alpha\beta}
   + N_{abcdef}(\sigma^{bcdef})_{\alpha\beta}
  \comma
\eqe
and similarly for $ (\hat{\alpha},\hat{\beta}) = (\alpha, \betabar),
 (\alphabar, \beta), (\alphabar,\betabar)$.
Consequently,  we have
\eqb
  \Delta^{2n} E_{i a} & \to & W_{ab} E_i^{ b} +
   S_{a-}(\theta \sigma^- D_i \theta)
  \comma \nn \\
   && \quad + \  M_{a-\ctil \dtil} (\theta \sigma^{-\ctil\dtil} D_i \theta)
   + N_{a-\ctil\dtil\tilde{e}\tilde{f}}
  (\theta \sigma^{-\ctil\dtil\tilde{e}\tilde{f}} D_i \theta) + \cdots
  \comma
\eqe
where the ellipses stand for similar terms obtained by replacing $\theta$
with $\thetabar$. Extracting the world-sheet derivative, we can apply the
lemma again. Since $W_{ab}, S_{a-}, M_{a-\ctil \dtil},
  N_{a-\ctil\dtil\tilde{e}\tilde{f}}$ have dimension 0,
we find that $W_{++} \to \calO (\theta^2)$;
$ W_{+-}, W_{\atil\btil},S_{+-}, M_{+-\ctil \dtil},
 N_{+-\ctil\dtil\tilde{e}\tilde{f}} \to \calO (\theta^0)$; and others are
vanishing. Thus, the only non-vanishing term among
$\Delta^{2n} E_{i a} \ (n \geq 1)$ is
$\Delta^2 E_{i+}  \sim \calO(\theta^2)$.

Similarly, for $\Delta^{2n-1} E_i^{ \alpha}$, we find that
\eqb
  \Delta^{2n-1} E_i^{ \alpha} &\to&
   W_a E_i^{ a} \theta^\alpha + S D_i \theta^\alpha + \cdots
  \period
\eqe
Since $W_a$ has dimension 1 and $S$ has 0, $W_+, S \to \calO(\theta^0)$
and $W_{-}, W_{\atil} \to 0$. Thus, $\Delta^{2n-1} E_i^{ \alpha}$
is non-vanishing only for $n=1$ with $\calO (\theta^1)$.

Summarizing, the terms of the forms in (\ref{higerterm}) are
at most of order two.
Therefore, we conclude that the light-cone Green-Schwarz action
in the plane-wave geometries with vanishing background fermions is
quadratic in the fermionic coordinates $\theta$ and $\thetabar$.

We also note that, in the proof, we have used only the properties
(I),(II') and (III). Thus, the proof holds for more general pp-wave
backgrounds as long as theses properties are satisfied.

\mysection{Explicit form of quadratic action in general backgrounds}
In the previous section, we have shown that the action (\ref{GSaction})
becomes quadratic in $\theta$ and $\thetabar$
for the plane- and pp-wave backgrounds which we are consider.  
Thus, we can obtain the complete action by performing
the $\theta$-expansion to second order. In this section,
we write down the explicit form of the action up to second order
in general bosonic on-shell backgrounds. 
The action for the backgrounds discussed in the previous section 
is obtained by substituting corresponding field configurations.

As we mentioned in Introduction,
the IIB Green-Schwarz action up to this order has been discussed
for general bosonic on-shell backgrounds: It has been derived
by starting from the membrane action in eleven dimensions
and by using T-duality in \cite{Cvetic-LPS}, or
by the $\theta$-expansion in \cite{Sahakian} (see also, \cite{Gates-MRV}).
For a class of plane-wave geometries
with vanishing dilaton and constant axion and other
field strengths, it has been written down in \cite{Metsaev-T2}.

Our motivations for discussing the action to quadratic order,
in spite of the existing results, are as follows. First, there are issues
on the precise form of the IIB Green-Schwarz action for general
backgrounds (see, e.g.,\cite{Corrado-HKW}). Thus,
for fixing the precise form, we think that it is useful
to compute it in a coherent manner described so far. Second,
it turns out that our results are different from those
in \cite{Sahakian} in signs. Third, we find that it is possible to
write down the action in a form in which its geometrical meaning is manifest
both in the Einstein and string frames.
In this way, the comparison becomes easier
with the action in \cite{Metsaev-T2} in corresponding cases,
which is used in recent works.

Recalling the first order terms in (\ref{L1}), we see that
the second order terms in $I^{(2)} = \half \Delta^2 I$
are given by
\eqb
  L^{(2)} 
   &\to&   
     \frac{1}{4} \Phi E_i^a
    \lb \sqrt{-g} g^{ij} \theta - \ep^{ij} \thetabar \rb \sigma_a
     \lbb \sigma_b (\Delta \Lambda) E_j^b
   - 2i \Delta \Ebar_j \rbb  + {\rm h.c.} \ \period \label{L21}
\eqe
By making use of the components of the superfields in Appendix B,
we find that 
\eqb
   \Delta E_j^\alpha &\to& E_j^a ( {\calD}_a \theta)^\alpha  \comma \quad
   \Delta \Lambda_\alpha \ \to \  ({\calD} \theta)_\alpha \comma \nn \\
   ({\calD}_a \theta)^\alpha
    &\equiv & D_a \theta^\alpha
   - i Z_{abcde}(\sigma^{bcde})^\alpha_{ \ \beta} \theta^\beta
   - \frac{3}{16} F_{abc} (\sigma^{bc})^\alpha_{ \ \beta} \thetabar^\beta
   + \frac{1}{48} F^{bcd} (\sigma_{abcd})^\alpha_{ \ \beta} \thetabar^\beta
   \comma \nn \\
   ( {\calD} \theta)_\alpha &\equiv&
    \frac{i}{2} P_a (\sigma^{a})^\alpha_{ \ \beta} \thetabar^\beta
    + \frac{i}{24} F_{abc} (\sigma^{abc})^\alpha_{ \ \beta} \theta^\beta
   \period \label{DeltaELambda}
\eqe
We note that $({\calD}_a \theta)^\alpha$ and
$ ({\calD}\theta)_\alpha$ are precisely the supersymmetry
transformations in the Einstein frame with parameter $\theta$
for the gravitino, $\delta_\theta \psi_a^\alpha$, and for
the spin $1/2$ field, $\delta_\theta \lambda_\alpha$, respectively
\cite{Howe-W}. We will see that
these terms add up to the supersymmetry transformation
of the gravitino in the string frame. This is consistent with the fact
that the passage from the Einstein frame to the string frame
involves mixing of the gravitino and spin $1/2$ field.
(In the above, all the superfields are evaluated at $z^M = z^M_0$.
Thus, if we faithfully follow the notation in \cite{Howe-W}, 
the fields should be denoted by small letters.  
However, here and in the following,
we often do not distinguish the capital and small letters,
since no confusion may occur.)
 
Substituting (\ref{DeltaELambda}) into (\ref{L21}) gives
\eqb
 L^{(2)} 
   &\to&   
      \frac{1}{4}  \Phi E_i^a E_j^b
    ( \sqrt{-g} g^{ij} \theta - \ep^{ij} \thetabar ) \sigma_a
    ( \sigma_b {\calD}\theta - 2i \overline{{\calD}_b \theta})
   +  {\rm h.c.} \ \period \label{L22}
\eqe
The term $(\sigma_b {\calD}\theta - 2i \overline{{\calD}_a \theta})$
is also rewritten in a form
$ -2i(\calD^{E;1}_b \thetabar + \calD^{E;2}_b \theta)$
with
\eqb
  \calD^{E;1}_b &=&  \del_b  - \frac{1}{4} \omega_{b,cd} \sigma^{cd}
   - \frac{1}{4} (\Pbar_b + \sigma_{bc} P^c)
   + i Z_{bcdef} \sigma^{cdef} \comma \nn \\
  \calD^{E;2}_b &=&
   \ - \frac{1}{48} (F^{cde} -\Fbar^{cde}) \sigma_{bcde}
     - \frac{1}{16} (F_{bcd} + 3\Fbar_{bcd}) \sigma^{cd}
   \period \label{DE}
\eqe
Here, the $U(1)$ connection in the covariant derivative
has been expressed by $P_a$ through (\ref{PQ}).

In sum, the IIB Green-Schwarz action in the Einstein frame
for general bosonic on-shell backgrounds is given by
\eqb
   I &=& \int d^2 \xi \, (L^{(0)} + L^{(2)}) \comma \nn \\
  L^{(0)} &=& \half \Bigl[ \sqrt{-g} g^{ij} \Phi E_i^{ a} (x) E_j^{ b}(x)
         \eta_{ab} + \epsilon^{ij} E^{b}_i (x) E^{a}_j (x)
   \calB_{ab} (x) \Bigr]
   \comma \\
  L^{(2)} &=& -\frac{i}{2}  \Phi E_i^{ a} E_j^{ b}
    ( \sqrt{-g} g^{ij} \theta - \ep^{ij} \thetabar ) \sigma_a
    ( \calD^{E;1}_b \thetabar + \calD^{E;2}_b \theta ) + {\rm h.c.}
   \comma \nn
\eqe
with $ \calB_{ab}(x)$ being the field $\calB_{ab}(z)$
evaluated at $z^M = z_0^M$,
and $\calD^{E;1(2)}_b $ given in (\ref{DE}).

We would also like to obtain the string-frame action. For this purpose,
we make a rescaling of the metric by $\Phi^{\half} = e^{\frac{\phi}{4}}$,
\eqb
   \tilde{e}_m^{\ a} =  e^{\frac{\phi}{4}} e_m^{\ a} \, ; \quad
   \sigma^a, X_{kl...}^{mn...} : \ {\rm fixed}
  \period
\eqe 
Under this transformation, for example,
$F_{abc} = e^{\frac{3}{4} \phi} \tilde{F}_{abc}$, and
\eqb
  E_j^{ a} \omega_{a,bc} \sigma^{bc} &=&
  \tilde{E}_j^{ a} ( \tilde{\omega}_{a,bc}
   + \eta_{ab} \tilde{\del}_c \ln \Phi )
   \sigma^{bc}
  \period
\eqe
By further rescaling the fermionic coordinates as
$ e^{\frac{\phi}{8}} \theta, e^{\frac{\phi}{8}} \thetabar
= \tilde{\theta}, \tilde{\thetabar}$, and dropping tildes, we then
arrive at the string-frame action
\eqb
  I &=& \int d^2 \xi \, (L_S^{(0)} + L_S^{(2)}) \comma \nn \\
  L_S^{(0)} &=& \half \Bigl[ \sqrt{-g} g^{ij} E_i^{ a} (x) E_j^{ b}(x)
         \eta_{ab} + \epsilon^{ij} E^{b}_i (x) E^{a}_j (x)
   \calB_{ab} (x) \Bigr]
   \comma \\
  L_S^{(2)} &=& -\frac{i}{2}  E_i^{ a} E_j^{ b}
    ( \sqrt{-g} g^{ij} \theta - \ep^{ij} \thetabar ) \sigma_a
    (  \calD^{S;1}_b \thetabar + \calD^{S;2}_b \theta ) + {\rm h.c.}
   \comma \nn
\eqe
with
\eqb
  \calD^{S;1}_b &=&  \del_b
     - \frac{1}{4} \omega_{b,cd} \sigma^{cd}
  - \frac{1}{4} [
     \Pbar_b + \sigma_{bc}(P^c + \half \del^c \phi) ]
   + i e^\phi Z_{bcdef} \sigma^{cdef} \comma \nn \\
  \calD^{S;2}_b &=&
   \ - \frac{1}{48} e^{\frac{\phi}{2}}(F^{cde} -\Fbar^{cde}) \sigma_{bcde}
     - \frac{1}{16} e^{\frac{\phi}{2}} (F_{bcd} + 3\Fbar_{bcd}) \sigma^{cd}
   \period \eqe

The quadratic term $L_S^{(2)}$ is written also in terms
of two Majorana-Weyl spinors $\vartheta^I \ (I = 1,2)$  defined by
\eqb
  \theta = \vartheta^1 + i \vartheta^2 \comma \quad
  \thetabar = \vartheta^1 - i \vartheta^2
 \period
\eqe
To proceed, we note that
\eqb
  \half (\theta A \thetabar + \thetabar \bar{A} \theta)
  &=& \vartheta^I [ \delta_{IJ} A^1 + (\rho_0)_{IJ} A^2 ]\vartheta^J
    \comma  \label{thetaA} \\
    \half (\thetabar A \thetabar + \theta \bar{A} \theta)
  &=&  \vartheta^I [ (\rho_3)_{IJ}  A^1
     + (\rho_1)_{IJ} A^2 ] \vartheta^J
    \ = \ \vartheta \rho_3 [A^1 + \rho_0 A^2] \vartheta
   \comma \nn
\eqe
where $A = A^1 + i A^2$ and
\eqb
  \rho_0 = \matrixii{0}{1}{-1}{0} \comma \quad
  \rho_1 = \matrixii{0}{1}{1}{0} \comma \quad
  \rho_3 = \matrixii{1}{0}{0}{-1} \period
\eqe
We have dropped indices $I,J$ in the last expression
in (\ref{thetaA}) in an obvious way.
From this, we find that any terms appear always in combinations
of the form $( \sqrt{-g} g^{ij} \vartheta - \ep^{ij} \vartheta \rho_3)$.
Collecting all the terms, we then have
\eqb
  L_S^{(2)} &=& - i  E_i^a E_j^b
    \lbb \sqrt{-g} g^{ij} \delta_{IJ} - \ep^{ij}(\rho_3)_{IJ} \rbb
   \vartheta^I \sigma_a \calD^S_b \vartheta^J
  \comma 
\eqe
with
\eqb
  \calD^S_b &=& \del_b - \frac{1}{4} \omega_{b,cd} \sigma^{cd}
     - \frac{1}{4} \Im{P^c} \rho_0 (\sigma_{bc} -\eta_{bc})
    + e^\phi Z_{bcdef} \rho_0 \sigma^{cdef} \nn \\
   && \quad 
     - \, \frac{1}{4} e^{\frac{\phi}{2}} \Re{F_{bcd}} \rho_3 \sigma^{cd}
       - \frac{1}{8} e^{\frac{\phi}{2}} \Im{F_{bcd}} \rho_1 \sigma^{cd}
     + \frac{1}{24} e^{\frac{\phi}{2}} \Im{F^{cde}} \rho_1 \sigma_{bcde}
  \period   
\eqe
Here, we have used (\ref{PQ}).

In order to further express this in terms of the fields in modern
string-theory conventions, we compare the equations of
motion in \cite{Howe-W}, whose notation we adopt in this paper, and those in
term of the fields in \cite{Polchinski} (with typos corrected):
\eqb
  R'_{ab} &=& -2 \Pbar_{(a} P_{b)} - \Fbar_{(a }^{ \ \, cd} F_{ b)cd}^{}
   + \frac{1}{12} \eta_{ab} \Fbar_{cde} F^{cde}
  - 384 Z_{a}^{\ cdef} Z_{bcdef} \comma \nn \\
 (R^{'P})_{ab} &=& 
  - \half { \del_{(a} \tau^P \del_{b)} \bar{\tau}^P \o (\Im{\tau^P})^2 }
  -\frac{1}{4} \lbb e^{-\phi^P} (H^P)_{acd} (H^P)_b^{\ cd}
   + e^{\phi^P} (\tilde{F}^P)_{acd} (\tilde{F}^P)_b^{\ cd} \rbb \\
  && \quad + \, \frac{1}{48} \eta_{ab}
     \lbb e^{-\phi^P} (H^P)_{cde} (H^P)^{cde}
   + e^{\phi^P} (\tilde{F}^P)_{cde} (\tilde{F}^P)^{cde} \rbb
   - \frac{1}{96} (\tilde{F}^P)_{acdef} (\tilde{F}^P)_b^{\ cdef}
  \comma \nn
\eqe
with $\tau^P = C^P_0 + ie^{-\phi^P}$.
We have put superscripts $P$ on the fields in \cite{Polchinski},
and the primes on the Ricci tensors; the sign of the
Ricci tensor in \cite{Polchinski} has also been flipped so as
to conform to the conventions in \cite{Howe-W}; the dilaton
 has been  denoted by $\phi^P$, to avoid confusion
with $\Phi$ in the action (\ref{GSaction}).
Comparing the above two equations gives
\eqb
   \phi &=& \phi^P \comma \quad
   P_a  \ = \ -\half (\del_a \phi^P + i e^{\phi^P} F^P_a ) \comma \nn \\
   F_{abc} &=& \half (e^{-\half \phi^P} H^P_{abc}
    + i e^{\half \phi^P} \tilde{F}^P_{abc}) \comma  \quad
   Z_{abcde}  \ = \ - \frac{1}{192} \tilde{F}^P_{abcde} \period
\eqe
To fix the signs and phases, we have used (\ref{PQ}) and 
other equations of motion. The relative signs of 
$F^P_a, H^P_{abc}, \tilde{F}^P_{abc}, \tilde{F}^P_{abcde}$
are, however, determined only up to the symmetry of the equations of motion.
Using these identifications and dropping superscripts $P$, we obtain
\eqb
  \calD^S_b &=& \del_b
  - \lb  \frac{1}{4} \omega_{b,cd} + \frac{1}{8} H_{bcd} \rho_3
\rb\sigma^{cd}
  \nn \\ 
  && \quad - e^{\phi} \lb \frac{1}{8} F^c \rho_0 \sigma_c
   + \frac{1}{48} \tilde{F}_{cde} \rho_1 \sigma^{cde}
   + \frac{1}{4 \cdot 480} \tilde{F}_{cdefg} \rho_0 \sigma^{cdefg} \rb
    \sigma_b     
  \period   
\eqe
This is in complete agreement with
the covariant derivative appearing in the string-frame supersymmetry
transformation of the gravitino, e.g., in \cite{Hassan}.
Note that different normalizations are also used in the literature:   
For example, the 5-form field strengths in \cite{Corrado-HKW} is different 
by a factor 4; the RR field strengths in \cite{Metsaev-T2} by a factor 2, 
which is absorbed into the shift of the constant part of dilaton.  
Thus, when $ \phi = F_a = 0$ and other field strengths are constant,
our quadratic terms in  $L^{(2)}$ agree with those in \cite{Metsaev-T2}.

\newpage
\mysection{Discussion}

In this paper, we have considered the IIB Green-Schwarz action in
general plane-wave backgrounds obtained as Penrose limits
from any IIB supergravity solutions with vanishing background fermions.
Starting from the Green-Schwarz action in terms of superfields
for general on-shell backgrounds, and using the $\theta$-expansion, 
we have shown that the action in
light-cone gauge is quadratic in the fermionic coordinates. 
For this to hold, the bosonic background fields do not have to be constant.
As long as the properties (I),(II') and (III) are satisfied,
the proof is valid also for more general pp-wave solutions,
of which the plane-wave solutions above are a subclass.
We have also discussed
the explicit form of the IIB Green-Schwarz action 
for general bosonic on-shell backgrounds up to second order
in the fermionic coordinates. We have written it down in a form
in which its geometrical meaning
becomes clear both in the Einstein and string frames.
This quadratic action (not necessarily in
light-cone gauge) is valid, up to this order, for any
on-shell backgrounds as long as background fermions are vanishing.
The complete action for the plane- and pp-waves 
which we have considered is read off
by substituting the corresponding background fields.
We have found that, when the dilaton and
1-form field strength are vanishing, and other field strengths are
constant, our string-frame action agrees with
the one in \cite{Metsaev-T2} up to conventions.
Our results may be used for further exploring strings in curved spacetime 
and RR backgrounds, string dualities, and the AdS/CFT correspondence.
They would also be of some use to fix the issues on the precise
form of the IIB Green-Schwarz action.

The proof for the IIB case may be easily applied to
other cases with appropriate modifications.
When the Yang-Mill part is set to be zero, the proof that the heterotic
Green-Schwarz action is quadratic in the fermionic coordinates is given
essentially just by truncating one Majorana-Weyl spinor, e.g.,
$\vartheta^2$. When the Yang-Mills part is turned on, the 3-form
field strength can have non-vanishing components for
$H_{\atil\btil\ctil} $ because of the Chern-Simons term and the
dimensionful coupling.
We may need appropriate changes in the analysis in this case.
A similar proof for the IIA case may also be given
essentially by flipping the chirality of one Majorana-Weyl spinor.
Although it is difficult to carry out
the $\theta$-expansion to higher orders in a totally general case,
the expansion would be under control
for special backgrounds such as maximally symmetric spaces.

\newpage
\begin{center}
  {\bf Acknowledgments}
\end{center}
We would like to thank S. Hyun, N. Ishibashi, Y. Kiem, K. Mohri,
J. Park, M. Sakaguchi and H. Shin
for useful discussions, conversations and correspondences.
Y.S. would also like to thank Korea Advanced Institute
of Science and Technology, where part of this work was done,
for its warm hospitality.
The work of S.M. is supported in part by Grant-in-Aid
for Scientific Research (C)(2) \#14540286 from
The Ministry of Education, Culture, Sports, Science
and Technology, whereas the work of Y.S. is supported in part
by University of Tsukuba Research Projects.

\vspace{7ex}
%
%
\ni {\large\bf Appendix}
\setcounter{section}{0}
\vskip -1ex
\appsection{Notation and conventions}
We use the following conventions for the indices:
\begin{center}
\begin{tabular}{lcl}
   $m,n, ... = 0,1,...,9 $ & & vector indices for spacetime    \\
   $\mtil ,\ntil, ... = 1,...,8 $
       & & vector indices for transverse directions of spacetime  \\
   $\mu,\nu,...; \, \bar{\mu}, \bar{\nu},...
    = 1,...,16  $ & & spinor indices for spacetime \\
   $M,N,... $ & & $M = (m,\mu,\bar{\mu}), \ N = (n,\nu,\bar{\nu}),...$ \\
   $a,b, ... = 0,1,..., 9 $ & & vector indices for tangent space \\
   $\atil,\btil, ... = 1,..., 8 $
    & & vector indices for transverse directions of tangent space \\
   $\alpha,\beta,...; \, \bar{\alpha}, \bar{\beta},...
    = 1,...,16  $ & & spinor indices for tangent space \\
   $\hat{\alpha}, \hat{\beta}, ...$ & &
    $\hat{\alpha} = (\alpha, \alphabar), \
   \hat{\beta} = (\beta, \betabar), ...$
  \\
  $A,B,... $ & &
    $A = (a,\alpha,\bar{\alpha}), \ B = (b,\beta,\bar{\beta}),...$
\end{tabular}
\end{center}

For the notation and conventions of the spinors, superspace and
super differential forms, we follow \cite{Howe-W}. For example,
$\eta_{ab} =$diag$(+1, -1, ..., -1)$, the scalar product of supervectors is
$U^A V_A = U^a V_a + U^\alpha V_\alpha - U^{\alphabar}V_{\alphabar}$,
the 5-form field strength is self-dual with $\ep^{01...9} = +1$, and so on.
We use the 16 component notation for the spinors. $\gamma$-matrices
are expressed by 16$\times$16 matrices $(\sigma^a)^{\alpha\beta}$ and
$(\sigma^a)_{\alpha\beta}$, which satisfy
$ \{ \sigma^a, \sigma^b\} = 2 \eta^{ab} $. The anti-symmetrization of
$\sigma^a$'s is, for example,
$\sigma^{ab} = \half (\sigma^a\sigma^b-\sigma^b\sigma^a)$. The Weyl
spinors with upper indices satisfy
$ (\sigma^0....\sigma^9)^\alpha_{\ \beta} \theta^\beta = - \theta^\alpha$.
Since the spin connection act on a vector $\varphi^a$ as
$D_m \varphi^a = \del_m \varphi^a + \varphi^b \omega_{m,b}^{\ \ \ a}
+ \cdots $,
the action on the spinors is determined as in (\ref{Dtheta}).

We often do not distinguish  the notation for the full superfields and
that for their lowest components in $\theta$,
since no confusion may occur.

\appsection{Components of superfields}
We list some components of the superfields which are derived
via constraints and Biachi identities. These are needed to calculate
the quadratic action in section 4 (and check the $\kappa$-symmetry).

\vskip 1ex
\ni
$T_{AB}^{\ \ \ C}$:
\eqb
  && T_{ab}^{\ \ c} = T_{\alpha\beta}^{\ \ c} = T_{\alpha b}^{\ \ c}
  = T_{\alpha\beta}^{\ \ \gamma} = T_{\alpha\betabar}^{\ \ \gammabar} = 0
   \comma  
   \nn \\
  && 
  T_{\alpha\betabar}^{\ \ c} = -i (\sigma^c)_{\alpha \beta} \comma \quad
  T_{a\beta}^{\ \ \gammabar}
   = -\frac{3}{16} \Fbar_{abc} (\sigma^{bc})_\beta^{\ \gamma}
     -\frac{1}{48} \Fbar^{bcd} (\sigma_{abcd})_\beta^{\ \gamma}
   \comma \\
   && T_{a\beta}^{\ \ \gamma}
   \to i Z_{abcde} (\sigma^{bcde})_\beta^{\ \gamma}
   \comma \quad 
  T_{a\betabar}^{\ \ \gammabar} = - (\overline{T_{a\beta}^{\ \ \gamma} })
   \comma \quad 
  T_{a\betabar}^{\ \ \gamma} = - (\overline{T_{a\beta}^{\ \ \gammabar} })
  \nn 
\eqe
where $\to$ stands for setting background fermions to be zero.

\vskip 1ex
\ni
$\calH_{ABC}$: \ in the gauge (\ref{gauge}),
\eqb
 && \calH_{\alpha\beta\gamma} = \calH_{\alpha\beta\gammabar} =
  \calH_{\alpha\betabar\gammabar} = \calH_{\alphabar\betabar\gammabar}
   = \calH_{a\beta\gammabar} = 0 \comma  \\
  && \calH_{a\beta\gamma} = \calH_{a\betabar\gammabar}
   = -i\Phi (\sigma_a)_{\beta\gamma} \comma  \quad
  \calH_{ab\gamma} = - \Phi (\sigma_{ab})_\gamma^{\ \delta}
    \bar{\Lambda}_\delta
  \comma \quad  
  \calH_{ab\gammabar} = - \Phi (\sigma_{ab})_\gamma^{\ \delta}
\Lambda_\delta
  \period \nn
\eqe

\ni
$ D \Lambda_\beta$:
\eqb
  D_\alpha \Lambda_\beta = -\frac{i}{24} F_{abc}
(\sigma^{abc})_{\alpha\beta}
  \comma \quad 
 \bar{D}_\alpha \Lambda_{\beta}
  = -\frac{i}{2} P_a (\sigma^{a})_{\alpha\beta}
\eqe
%
%
\def\thebibliography#1{\list
 {[\arabic{enumi}]}{\settowidth\labelwidth{[#1]}\leftmargin\labelwidth
  \advance\leftmargin\labelsep
  \usecounter{enumi}}
  \def\newblock{\hskip .11em plus .33em minus .07em}
  \sloppy\clubpenalty4000\widowpenalty4000
  \sfcode`\.=1000\relax}
 \let\endthebibliography=\endlist
%
%
\vskip 10ex
\begin{center}
 {\bf References}
\end{center}
\par \smallskip

\end{document}